\documentclass[nofootinbib,prd,twocolumn,showpacs,showkeys,preprintnumbers]{revtex4-1}
\usepackage{hyperref,amssymb,amsmath,mathrsfs,bm,graphicx}
\begin{document}
\title {REVERSIBLE DISSIPATIVE PROCESSES, CONFORMAL MOTIONS AND LANDAU DAMPING}
\author{L. Herrera} 
\email{laherrera@cantv.net.ve}
\altaffiliation{Also at U.C.V., Caracas}
\author{A. Di Prisco}
\email{adiprisc@fisica.ciens.ucv.ve}
\altaffiliation{Also at U.C.V., Caracas}
\author{J. Ib\'a\~nez  } 
\email{j.ibanez@ehu.es}
\affiliation{Departamento de F\'\i sica Te\'orica e Historia de la Ciencia,
Universidad del Pa\'{\i}s Vasco, Bilbao, Spain}
\date{\today}
\begin{abstract}
The existence of a dissipative flux vector is known to be compatible with reversible processes, provided a timelike conformal Killing vector  (CKV) $\chi^\alpha=\frac{V^\alpha}{T}$ (where $V^\alpha$  and $T$ denote the four--velocity and temperature respectively) is admitted by the space--time. Here we show that if  a constitutive transport equation, either within the context of standard irreversible thermodynamics or  the causal Israel--Stewart theory, is adopted, then such a compatibility also requires vanishing dissipative fluxes. Therefore, in this later case  the vanishing of  entropy production generated by the existence of such CKV is not actually associated to an imperfect fluid, but to a non--dissipative one. We discuss also about  Landau damping.
\end{abstract}
\date{\today}
\pacs{04.40.-b, 04.40.Nr, 05.70.Ln}
\keywords{Relativistic fluids, dissipative fluids, causal dissipative theories.}
\maketitle
\section{Reversibility and CKV}
In a recent work \cite{1B} it was shown that   Lemaitre--Tolman--Bondi spacetimes (LTB) as seen by a tilted observer are  real dissipative (i.e. irreversible). This conclusion being derived from the symmetry properties of such spacetimes.  

Indeed, it is known  that, in principle, 
 imperfect fluids are not necessarily incompatible with reversible processes  (e.g see \cite{14}--\cite{16}).
In the context of the standard Eckart theory \cite{17} a sufficient condition for the compatibility of an imperfect fluid with vanishing entropy production (in the absence of bulk viscosity) is  the existence of a conformal Killing vector field CKV) $\chi^\alpha$ such that $\chi^\alpha =\frac{V^\alpha}{T}$ where $V^\alpha$ is the four--velocity of the fluid and $T$ denotes the temperature. In the context of   causal  dissipative  theories, e.g. \cite{18}--\cite{23}, as shown in \cite{1B}, the existence of such CKV  also appears as a condition for an imperfect fluid to be compatible with vanishing entropy production. However,  as is known  LTB does not admit CKV \cite{hltb}.

The purpose of this note is to show  that the compatibility of reversible processes and  the existence of dissipative fluxes becomes trivial if a constitutive transport equation is adopted, since in this later case such compatibility  forces the heat flux vector to vanish as well. In other words, even if {\it ab initio} the fluid is assumed imperfect (non--vanishing heat flow vector) the imposition of the CKV and the vanishing entropy production condition may cancel the heat flux,  depending on the transport equation assumed.

The proof of the abovementioned statement follows from a theorem about timelike CKV shown by Oliver and Davis \cite{th1} (see also \cite{th2}).

Indeed it has been shown that a  space--time admits a timelike CKV (a conformal motion)  $\chi^\alpha=\chi V^\alpha$, i.e. 
\begin{equation}
\mathcal{L}_\chi g_{\alpha \beta} =\psi g_{\alpha \beta},
\label{1}
\end{equation}
 where $\mathcal{L}_\chi$ denotes the Lie derivative with respect to the vector field ${\bold \chi}$, if and only if:
\begin{equation}
\sigma_{\alpha \beta}=0,
\label{2}
\end{equation}
and 
\begin{equation}
 a_{\alpha}=(\log \chi)_{,\alpha}+\frac{\Theta}{3}V_\alpha,
\label{3}
\end{equation}
furthermore, it turns  out that $\psi=\frac{2}{3}\chi \Theta$, where $\sigma_{\alpha \beta}$, $a_{\alpha}$ and $\Theta$ denote the shear, the four--acceleration and the expansion of the fluid, respectively.

Next, the transport equation for the Israel--Stewart theory reads
\begin{equation}
\tau h^{\alpha \beta}V^\gamma q_{\beta;\gamma}+q^\alpha=-\kappa h^{\alpha \beta}\left(T_{,\beta}+Ta_\beta\right)-\frac{1}{2}\kappa T^2 \left(\frac{\tau V^\beta}{\kappa T^2}\right)_{;\beta} q^\alpha,
\label{4}
\end{equation}
where $\kappa$ denotes the thermal conductivity,  $T$ and $\tau$ denote temperature and relaxation time respectively, $h_{\alpha \beta}$ is the projector onto the hypersurface orthogonal to $V^\alpha$,  and  $q^\mu$ is the heat flux vector.  If $\tau=0$ we recover the well known Eckart--Landau equation.

In the spherically symmetric case under consideration, the transport equation has only one independent component which may be obtained from (\ref{4}) by contracting with the unit spacelike vector $s^\alpha$ orthogonal to $V^\alpha$, we get
\begin{equation}
\tau V^\alpha  q_{,\alpha}+q=-\kappa \left(s^\alpha T_{,\alpha}+T a\right)-\frac{1}{2}\kappa T^2\left(\frac{\tau V^\alpha}{\kappa T^2}\right)_{;\alpha} q,
\label{5}
\end{equation}
where $q^\alpha=q s^\alpha$ and $a^\alpha=a s^\alpha$.

On the other hand, putting $\chi=\frac{1}{T}$, we obtain from (\ref{3})
\begin{equation}
Ta+s^\alpha T_{,\alpha}=0,
\label{6}
\end{equation}
which combined with (\ref{5}) produces
\begin{equation}
\tau V^\alpha  q_{,\alpha}+q=-\frac{1}{2}\kappa T^2\left(\frac{\tau V^\alpha}{\kappa T^2}\right)_{;\alpha} q.
\label{7}
\end{equation}

Now, in \cite{1B} it was shown that from the Gibbs equation and Bianchi identities, it follows 
\begin{equation}
S^{\alpha}_{;\alpha} = -\frac{1}{2} T^{\alpha \beta}_{dis.}
\mathcal{L}_\chi g_{\alpha \beta} -\frac{1}{2}\left(\frac{q^2V^\mu \tau}{\kappa T^2}\right)_{;\mu},
\label{diventropiaIV}
\end{equation}
where $S^\alpha$ is the entropy four--current,   and  $T^{\alpha \beta}_{dis.}=V^\alpha q^\beta+V^\beta q^\alpha$. 

From the above it follows that for a generic reversible process (vanishing production of entropy) we should have that both terms on the right of (\ref{diventropiaIV})  vanish simultaneously. Thus if (\ref{1}) is admitted we also should demand (see \cite{1B} for a detailed discussion)
\begin{equation}
\left(\frac{q^2V^\mu \tau}{\kappa T^2}\right)_{;\mu}=0.
\label{8}
\end{equation}

Then combining (\ref{8}) with (\ref{7}) one obtains at once
\begin{equation}
q=0.
\label{9}
\end{equation}

In other words, in the presence of a CKV of the kind considered here, the  assumption of a  transport equation such as (\ref{4}) implies that  a vanishing entropy production leads to a vanishing heat flux vector.Therefore, under the conditions above   the system is not only reversible but also not dissipative. Nevertheless, it is worth mentioning that dissipative reversible processes do exist. We shall briefly comment about one known example in the next section.

\section{Landau damping}
If no assumption is made about the transport equation,  reversible dissipative processes may occur. An example of which is  the well known Landau damping  in collisionless plasma \cite{72}, \cite{73}, whose existence has been confirmed by experiments\cite{LD1}, \cite{LD2}. In that  case, the dissipation is related to electrons whose speed in the direction of propagation of an electric wave, equals the phase speed of  the latter. It happens then that such an electric field is stationary with respect to those electrons and therefore the time average of the work done on the later does not vanish. The argument to infer the reversibility is based on the fact that the plasma is collisionless. 

However,  it has been shown  that in some cases, forces not related to binary interactions may be interpreted  as collisional terms appearing in the Boltzmann equations, and thereby producing entropy  \cite{74}. Basically, what authors of \cite{74} show is that a specific collisional interaction may be mapped onto an effective force, implying thereby that there exists a certain freedom to interpret collisional events (producing entropy) in terms of forces (and viceversa). On the basis of this argument one might speculate about the possibility that  the force exerted by  the electric wave could be interpreted as a collisional term, producing an irreversible process.  This however seems not to be the case as it follows from the fact  that Landau damping can be described by a Hamiltonian formulation thus demonstrating its conservative character \cite{manu}.

\section{Summary}
If a timelike CKV of the kind considered above is assumed to exist, then in principle  a nonvanishing heat flux vector may be compatible with a reversible process. However, as soon as a general transport equation as (\ref{4}) is adopted, the CKV and the condition of reversibility produce  the  vanishing of the  heat flux vector and the system is no longer dissipative.

An example of  a ``truly''  dissipative reversible process is provided by Landau damping. Of course, for a phenomenon  taking place in collisionless plasma, it is not reasonable to assume   a transport equation as (\ref{4}).

Before concluding, the following remark is in order: for many physicists the expression ``reversible dissipative process'' may sound contradictory, and as a matter of fact for some of them   Landau damping is not a ``truly'' dissipative process since it is reversible (see for example \cite{r1}, \cite{r2}). Here  we have used    the term ``dissipative'' to  mean non--vanishing of the heat flux, independently on whether or not  entropy is increasing, in the same spirit as  in \cite{14}--\cite{16}.

\begin{acknowledgments}
LH wishes to thank Manuel Valle for helpful discussions about Landau damping, and  Fundaci\'on Empresas Polar for financial support and Departamento   de F\'{\i}sica Te\'orica e Historia de la  Ciencia, Universidad del Pa\'{\i}s Vasco, for financial support and hospitality. ADP  acknowledges hospitality of the
Departamento   de F\'{\i}sica Te\'orica e Historia de la  Ciencia,
Universidad del Pa\'{\i}s Vasco. This work was partially supported by the Spanish Ministry of Science and Innovation (grant FIS2010-15492). 
\end{acknowledgments}


\begin{thebibliography}{100}

\bibitem{1B} L. Herrera, A. Di Prisco and J. Ib\'a\~nez {\it Phys. Rev. D} {\bf 84}, 064036 (2011).
\bibitem{14} H. Stephani {\it Introduction to General Relativity} (Cambridge: Cambridge University Press) (1982).
\bibitem{15} M. L. Bedran and M. O. Calvao {\it Class. Quantum Grav.} {\bf 10}, 767 (1993).
\bibitem{16}  J. Triginer and D. Pav\'on, {\it Class. Quantum. Grav.} {\bf 12}, 199 (1995).
\bibitem{17} C. Eckart {\it Phys. Rev.} {\bf 58}, 919 (1940).

\bibitem{18}  I. M\"{u}ller {\it Z. Physik} {\bf 198}, 329 (1967).
\bibitem{19} W.  Israel {\it Ann. Phys.} (NY) {\bf 100}, 310 (1976).
\bibitem{20} W. Israel and J. Stewart {\it Phys. Lett. A} {\bf 58}.
213  (1976).
\bibitem{21} W. Israel and J. Stewart {\it Ann. Phys.} (NY) {\bf 118}, 341 (1979).
\bibitem{22} D. Jou, J. Casas-V\'azquez and G. Lebon {\it Rep. Prog. Phys.}
{\bf 51}, 1105 (1988).
\bibitem{23}  D. Jou, J. Casas-V\'azquez and G. Lebon {\it Extended irreversible Thermodynamics}
(Berlin: Springer)  (1993).
\bibitem{hltb} L. Herrera, A. Di Prisco, J. Ospino  and J. Carot {\it Phys. Rev. D} {\bf 82}, 024021 (2010).
\bibitem{th1} D. R. Oliver, Jr.  and  W. R. Davis {\it Gen.Rel. Grav.} {\bf 8}, 905 (1977).
\bibitem{th2} K. L. Duggal {\it J. Math. Phys.} {\bf 28}, 2700 (1987).
\bibitem{72} L. Landau {\it J. Phys. (U.S.S.R.)} {\bf 10}, 25 (1946).
\bibitem{73} E. Lifchitz and L. Pitayevski {\it Cin\'etique Physique} (Ed. Mir, Moscow), (1990).
\bibitem{LD1} J. H. Malmberg and C. B. Wharton {\it Phys. Rev. Lett.} {\bf 13}, 184 (1964).
\bibitem{LD2} H. Derfler and T. C. Simonen {\it J. Appl. Phys.} {\bf 38}, 5014 (1967).
\bibitem{74} W. Zimdahl, D. J. Schwarz, A. B. Balakin and D. Pav\'on {\it Phys. Rev. D} {\bf 64}, 063501 (2001).
\bibitem{manu} M. Valle {\it Phys. Lett. A} {\bf 375}, 666 (2011).
\bibitem{r1} Z. Sedlacek {\it AIP Conf. Proc.} {\bf 345},119 (1995).
\bibitem{r2} ZG Bertaina, L. Pitaevskii and S. Stringari {\it Phys. rev. Lett.} {\bf 105}, 150402  (2010).
\end{thebibliography}
\end{document}